\documentclass[a4paper,twocolumn,11pt]{quantumarticle}
\pdfoutput=1
\usepackage[utf8]{inputenc}
\usepackage[english]{babel}
\usepackage[T1]{fontenc}
\usepackage{amsmath}
\usepackage{hyperref}

\usepackage[utf8]{inputenc}
\usepackage[english]{babel}
\usepackage[T1]{fontenc}
\usepackage{amsmath}
\usepackage{hyperref}

\usepackage{xfrac}
\usepackage{amsfonts}
\usepackage{dsfont}
\usepackage{amsmath}
\usepackage{amssymb}
\usepackage{mathtools}
\usepackage{graphicx}
\usepackage{color}
\usepackage{bbm}
\usepackage[normalem]{ulem}

\usepackage[numbers, sort&compress]{natbib}

\newcommand{\ignore}[1]{}

\newcommand{\ket}[1]{|#1\rangle}
\newcommand{\bra}[1]{\langle #1|}
\newcommand{\ketbra}[2]{|#1\rangle \hspace{-2pt}\langle #2|}

\newcommand{\Andreu}[1]{{\color{red}ARC:#1}}

\DeclareMathAlphabet{\mathcal}{OMS}{cmsy}{m}{n}

\begin{document}

\title{Measurement-induced resetting in open quantum systems}

\author{Andreu Riera-Campeny}
\affiliation{F\'{\i}sica Te\`{o}rica: Informaci\'{o} i Fen\`{o}mens Qu\`{a}ntics. Departament de F\'{\i}sica, Universitat Aut\`{o}noma de Barcelona, 08193 Bellaterra, Spain}
\orcid{0000-0003-3260-993X}

\author{Jan Oll\'e}
\affiliation{Institut de F\'{\i}sica d'Altes Energies (IFAE). The Barcelona Institute of Science and Technology (BIST). Universitat Aut\`{o}noma de Barcelona, 08193 Bellaterra, Spain}
\orcid{0000-0003-3338-5130}

\author{Axel Mas\'o-Puigdellosas}
\affiliation{F\'{\i}sica Te\`{o}rica: Grup de Física Estadística. Departament de F\'{\i}sica, Universitat Aut\`{o}noma de Barcelona, 08193 Bellaterra, Spain}


\begin{abstract}
We put forward a novel approach to study the evolution of an arbitrary open quantum system under a resetting process. Using the framework of renewal equations, we find a universal behavior for the mean return time that goes beyond unitary dynamics and Markovian measurements. Our results show a non-trivial behavior of the mean switching times with the mean measurement time $\tau$, which permits tuning $\tau$ for minimizing the mean transition time between states. We complement our results with a numerical analysis, which we benchmark against the corresponding analytical study for low dimensional systems under unitary \textit{and} open system dynamics. 
\end{abstract}	

\section{Introduction.}

A \textit{reset} is a process that brings a system in any given state to a normal condition or reset state. Resets might be beneficial to the user since they allow for a safe fresh start under controlled conditions. This conceptually simple idea appears in many areas of physics from information management protocols to the modelling of animal search processes. For example, a computer can use a reset to avoid crashing and a bird that recurrently comes back to its nest enhances the rate of finding hidden food.

Classical stochastic processes \textit{with} resets have attracted a lot of attention lately. It has been shown that the inclusion of resetting can be beneficial for many practical purposes. For instance, the first passage of a diffusive random walker under Markovian resetting has been shown to attain an optimum value for a finite reset rate \cite{EvMa11}. This optimization capacity has been observed for different types of walkers and resetting mechanisms \cite{EvMa11p,MoVi13,ChSc15,CaMe15,PaKu16,NaGu16,MoMa17,Sh17,MaCaMe19,KuGu19,Ma19,BoChSo19,SaInAj19,MaMo19,Si20}, even when a time penalty is considered after the resets \cite{EvMa19, MaCaMe19_1, PaKuRe19}, for enzymatic inhibition \cite{ReUrKl14,RoReUr15} or the completion of a Bernoulli trial process \cite{Be18}. Also, general analyses of the completion time of stochastic processes under resets have been done in \cite{PaRe17,Reuveni16,ChSo18,DeRaRe20}. An in-depth review on stochastic resetting can be found in \cite{EvMaSc20}.

In the quantum world, a Markovian master equation describing the evolution of a system undergoing unitary dynamics with a fixed reset state was introduced for the first time in \cite{HaDuBr2006}. There, the authors studied the prevalence of entanglement in open quantum systems. The same \textit{reset} master equation has been used further as a model for open quantum dynamics \cite{LiPoSk2010, TaHaBrBrBo2020}. Recently, the spectral properties of the generator of the reset master equation as well as the connection between classical and quantum reset processes have been investigated \cite{RoToLeGa2018}. Moreover, equivalent results can also be found studying a Markovian resetting process added on top of the otherwise unitary dynamics, without the explicit use of the reset master equation \cite{MuSeMa18}.

Despite the remarkable progress done in the aforementioned works, it is not clear under which circumstances such resetting dynamics would arise. Here, we consider an inherent resetting process built already in the postulates of quantum mechanics: the measurement-induced \textit{collapse} of the wave-function. This collapse occurs when an external agent performs a measurement of the quantum state inducing its decoherence. Then, the \textit{post-measurement} (or collapsed) state can be interpreted as a reset state of the dynamics. 

It is worth mentioning that the type of dynamics we propose in this work resemble the arising in the so-called quantum walks \cite{AhDaZa93,Me96,Ke03,GrVeWeWe13}. There, a quantum system under unitary evolution is recurrently measured, often stroboscopically, and the overall dynamics are analyzed. Particularly, the probability of reaching a target state \cite{ThBaKe18,ThMuKeBa20} and the time needed to do so \cite{FrKeBa16,FrKeBa17,YiZiThBa19} are the studied magnitudes. 

Nevertheless, there are several open questions that require further research. For instance: (i) Is it possible to go beyond the stroboscopic and the Markovian regime? (ii) Is it possible to go beyond the unitary dynamics paradigm? (iii) Does resetting display any universal behavior? (iv) When is stochastic quantum resetting beneficial? In this work, we put forward a formalism based on renewal equations in order to tackle questions (i)--(iv).  

\section{Set up.}

We consider an $N$-dimensional, open quantum system at a given initial state $\rho(0) = \rho_0$, whose evolution is formally described by a completely-positive and trace-preserving map $\rho(t) = \mathcal{E}(t)[\rho_0]$, which we call for simplicity a quantum evolution $\mathcal{E}(t)$. The most general form of a quantum evolution is given by the Kraus decomposition \cite{NiCh2002}
\begin{equation}
    \mathcal{E}(t)[\rho] = \sum_k \text{A}_k(t) \rho\text{A}_k(t)^\dagger,\label{eq:kraus_decomposition}
\end{equation}
where $\text{A}_k$ are the Kraus operators and fulfill the completeness relation $\sum_k \text{A}_k(t)^\dagger \text{A}_k(t) = \mathbf{1}$ at all times $t$. The simplest case of quantum evolution corresponds to having a single Kraus operator $\text{A}_1(t)$. In that case, the completeness relation implies that the evolution is unitary. From now on, we restrict ourselves to time-homogeneous evolutions $\mathcal{E}(t+s) = \mathcal{E}(t)\mathcal{E}(s)$ for the sake of the discussion. 

Additionally, we consider that the evolution $\mathcal{E}(t)$ can be interrupted at stochastic times $t=t_1,t_2,\cdots$ in which a measurement $\text{M}$ is performed. At this point, the dynamics are restarted, being the outcome of the measurement the new initial state. The stochastic times $t=t_1,t_2,\cdots$ are such that the differences $\Delta t_i=t_k-t_{k-1}$ are sampled from a given probability density function $\varphi(t)$, the \textit{measurement time} distribution. The outputs $m_i$ for $i=1,\cdots,N$ of the measurement $\text{M}$ are assumed to be non-degenerate, and the corresponding eigenstates are denoted by $|m_i\rangle$.

\begin{figure}[t] 
\centering
\includegraphics[width=0.45\textwidth]{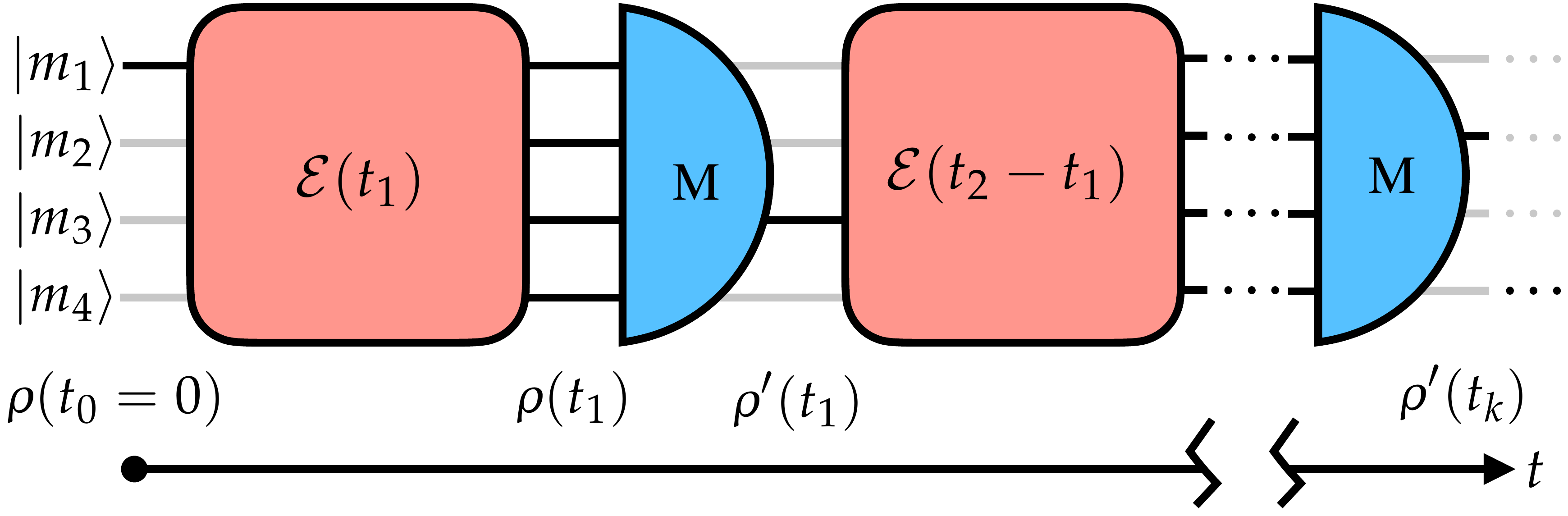}
\caption{Sketch of the evolution considered for $N=4$, in which the evolution $\mathcal{E}(t)$ starting from $\ket{m_1}$ is interrupted by the measurement $\text{M}$. The black lines represent the "trajectory" of the state during the evolution. After each measurement, the state is reset to one of the states $\ket{m_i}$.}
\label{fig:sketch}
\end{figure}

We consider the following process: at time $t_0 = 0$ the system is in the initial state $\rho_0 = \ketbra{m_i}{m_i}$ where $i =1, 2,\cdots, N$. We select a target state $i=\star$ for further convenience (which, without loss of generality, can be thought as $\ket{m_\star} =\ket{m_1}$). Then, the system evolves up to time $t_1$ to the state $\rho(t_1) = \mathcal{E}(t_1)[\rho_0]$. Subsequently, the measurement $\text{M}$ is performed yielding the collapsed state $\rho_1'$ corresponding to one of the pure states $\ket{m_i}$. This process is then iterated to obtain the state $\rho(t_{k}) = \mathcal{E}(\Delta t_{k})[\rho'(t_{k-1})]$ at time $t_k$. This process is sketched in Fig.~\ref{fig:sketch}. 

The ``experimentally'' accessible information of the aforementioned evolution process is encoded in the probabilities 
\begin{equation}
    p(i,t|j) = \langle m_i|\mathcal{E}(t)[\ketbra{m_j}{m_j}]|m_i\rangle. 
\end{equation}
However, tracking all the $p(i,t|j)$ at each instant of time (i.e., locally in time) requires a large amount of resources. Instead, it would be more efficient to obtain information about global properties of the evolution. This global information is gathered, for instance, in the mean switching times $T_i$, which are defined as the average time of jumping from any state $\ket{m_i}$ to the target state $\ket{m_\star}$. The set of $N-1$ mean switching times is completed by incorporating the mean return time $T_\star$, i.e. the average time of a closed loop that starts and ends in the target state $\ket{m_\star}$.

The mean return time $T_\star$ and the mean switching times $\{T_i\}$ are random variables themselves. We denote its probability density function $q_\star(t)$ and $\{q_i(t)\}$ respectively. Being $Q_\star(t)=\int_t^\infty q_\star(t')dt'$ and $Q_i(t)=\int_t^\infty q_i(t')dt'$ the survival probabilities of the probability density functions, the return times can be computed as
\begin{align}
&T_\star = \int_0^\infty t q_\star(t) dt=\hat{Q}_\star(0),\label{eq:Testar}\\    
&T_i =  \int_0^\infty t q_i(t) dt=\hat{Q}_i(0),\label{eq:Tei}
\end{align}
where the ``hat'' denotes the Laplace transform, i.e. $\hat{f}(s) = \mathbb{L}[f(t)](s) = \int_0^\infty dt f(t) e^{-s t}$ for any time-dependent function $f(t)$. 

The survival probability $Q_\star(t)$ (resp. $Q_i(t)$) is defined as the probability of \textit{not} having measured the outcome $m_\star$ (resp. $m_i$) after time $t$, and obey the following renewal equations
\begin{align}
&Q_\star(t) = \bar{\varphi}(t) + \sum_{j\neq\star} \int_0^t \varphi(t') p(j,t'|\star) Q_j(t-t')dt',\label{eq:renewalstar}\\
&Q_i(t) = \bar{\varphi}(t) + \sum_{j\neq\star}\int_0^t \varphi(t') p(j,t'|i) Q_j(t-t')dt',\label{eq:renewal}
\end{align}
where $\bar{\varphi}(t)=\int_t^\infty \varphi(t') dt'$ denotes the probability of not having performed any measurement at time $t$. Equation~\eqref{eq:renewalstar} (and similarly Eq.~\eqref{eq:renewal}) reads as follows: the probability of not having detected the outcome $\star$ at time $t$ is the probability of not having measured the system plus the probability of having measured the system at least once at a time $t'<t$, obtained an outcome $j \neq \star$, from which, with probability $Q_j(t-t')$, the outcome $\star$ has not yet been detected at time $t$. The system of integral Eqs.~\eqref{eq:renewalstar}--\eqref{eq:renewal} can be solved by means of a Laplace transform. In the following, we provide the analytic solution and derive our main results. 

\section{Results}

After applying the Laplace transform at both sides of the equalities, the system of Eqs.~\eqref{eq:renewalstar}--\eqref{eq:renewal} becomes algebraic. Hence, it is convenient to introduce the matrix $\textbf{W}_{ij}(s) = \mathbb{L}[\varphi (t)p(i,t|j)](s)$ with $i,j \neq \star$, and the vector $\textbf{w}_{\star_i} = \mathbb{L}[\varphi (t)p(i,t|\star)](s)$. Similarly, one can gather in the vector $\textbf{T}$ all the mean switching times $T_i$ for $i\neq \star$. Finally, using Eqs.~\eqref{eq:Testar}--\eqref{eq:Tei} one arrives at
\begin{align}
    &T_\star = \tau(1+ \textbf{w}_{\star}^T(0)( \mathbf{1}-\mathbf{W}^T(0))^{-1}\textbf{v}_1) ,\label{eq:tstar}\\
    &\textbf{T} = \tau( \mathbf{1}-\mathbf{W}^T(0))^{-1}\textbf{v}_1,\label{eq:tis}
\end{align}
where $\tau = \int_0^\infty t \varphi(t) dt$ and $\textbf{v}_1=(1,\cdots,1)$ is a vector of ones of length $N-1$.  Equations~\eqref{eq:tstar}--\eqref{eq:tis} constitute our first main result. We refer the interested reader to Appendix \ref{appendix:derivation_of_eqs_7_and_8} for details on the derivation. A slightly different renewal approach on this topic has been recently formulated in \cite{FrKeBa16} to study unitary evolution with stroboscopic measurements. 

Equations~\eqref{eq:tstar}--\eqref{eq:tis} represent a compact form of the mean return and mean switching time of \textit{any} quantum evolution $\mathcal{E}(t)$. However, they are as general as uneasy to interpret. Some physical intuition
can be obtained by restricting the form of quantum evolution $\mathcal{E}(t)$ under consideration.

A natural starting point consists in considering the so-called \textit{unital} quantum evolutions (analogous to classical processes with doubly stochastic matrices). A completely-positive and trace-preserving map represents a unital quantum evolution if preserves the identity, i.e. $\mathcal{E}(t)[\textbf{1}] = \textbf{1}$, which implies that the maximally mixed state is an equilibrium state. The most simple example of unital evolution corresponds to the unitary evolution $\mathcal{E}(t)[\rho] = \text{U}(t)\rho\text{U}(t)^\dagger$. For a unital quantum evolution one has that $\sum_j p(i,t|j) = 1$. Then, it is possible to derive the closure relation
\begin{equation}
    \textbf{w}_\star(0) + \textbf{W}(0)\mathbf{v}_1 = \mathbf{v}_1\label{eq:closure_relation}.
\end{equation}
See Appendix \ref{appendix:derivation_of_eq_9} for details. Finally, using Eq.~\eqref{eq:closure_relation} into Eq.~\eqref{eq:tstar}, one finds the universal result
\begin{equation}
    T_\star = N\tau,\label{eq:Ntau}
\end{equation}
which holds for \textit{any} unital quantum evolution. In particular, any closed quantum system fulfills Eq.~\eqref{eq:Ntau}. 

The universal result in Eq.~\eqref{eq:Ntau} may seem unintuitive at first glance. For instance, consider a system undergoing a periodic evolution with period $\tau_\star$ such that $\mathcal{E}(\tau_\star)[\ketbra{m_\star}{m_\star}] = \ketbra{m_\star}{m_\star}$. Then, using a deterministic measurement time distribution $\varphi(t) = \delta(t-\tau_\star)$ leads trivially to $T_\star = \tau_\star$. Nevertheless, any finite amount of noise in either $\varphi(t)$ or in $\mathcal{E}(t)$ would bring us back to Eq.~\eqref{eq:Ntau}. Mathematically, those non well-behaved cases correspond to singularities of the matrix $\mathbf{1}-\mathbf{W}^T(0)$. 

In particular, care has to be taken when applying Eq.~\eqref{eq:Ntau} in the presence of conservation laws or, more precisely, when the unital quantum evolution $\mathcal{E}(t)$ has a block structure. To illustrate this issue, consider the case where the eigenstates $\ket{m_i}$ are either dynamically connected or disconnected from the target state $\ket{m_\star}$. Denoting by $N_c$ the number of connected states, we find the modified universal behavior
\begin{equation}
    T_\star = N_c \tau,
    \label{eq:NcTau}
\end{equation}
valid for unital quantum dynamics in the presence of conservation laws. We refer the interested reader to Appendix \ref{appendix:universal_behavior_under_block_structure} for an extended discussion. Thus, for unital evolutions, $T_\star$ can be computed by simply multiplying the mean measurement time $\tau$ by the mean number of possible outcomes $N_c$ before detecting $\ket{m_\star}$. This property for the product of random variables is found under the name of Wald's identity in the mathematical bibliography \cite{Wa44,Wa45}. Using a different approach, Eq.\eqref{eq:NcTau} is derived in \cite{GrVeWeWe13} for unitary evolution and deterministic measurement times, \textit{i.e.}  $\varphi(t)=\delta(t-\tau)$. Similar laws are known in the context of queuing theory for general distributions of the random variables \cite{Li08}.

At this point, we see that there is a universal behaviour in the mean return time when the evolution is unital, while the mean first switching times depend strongly on the evolution itself. To illustrate these different behaviors, we analytically solve three different types of dynamics for a qubit system.

\begin{figure*}
\includegraphics[width=\textwidth]{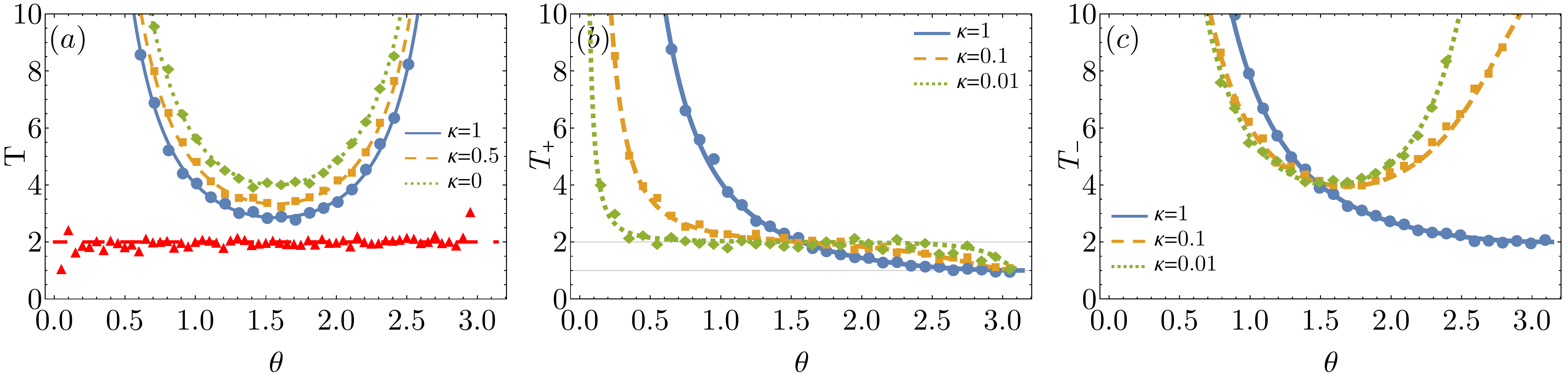}
\caption{Comparison between the analytic and numerical simulation results. Every data point has been obtained by simulating the quantum process in Fig.\ref{fig:sketch} 1,000 times. Units are such that $\omega = \tau^{-1} = 1 $, while $\kappa$ is varied.  (a) Results for the qubit with unitary and unital dynamics. Lines correspond to analytic predictions, while markers are results from simulations. The solid (blue), dashed (yellow) and dotted (green) lines are switching times $T_-$, while the dot-dashed (red) line is $T_+$. The data for $T_+$ (triangle markers) contains both $\kappa = 0$ and $\kappa = 1$ results from the simulations. (b) Results for $T_+$ in the open qubit system with non-unital dynamics. (c) Results for the switching time $T_-$ in the open qubit system with non-unital dynamics.}\label{fig:all}
\end{figure*}

\subsection{The qubit (N=2)}

A paradigmatic example of closed and open quantum evolution is the qubit. Here, we analyze this two dimensional system to provide a comprehensive example of the results derived above. The three cases that we consider are: (i) unitary dynamics, (ii) unital (non-unitary) dynamics implemented by \textit{dephasing noise}, and (iii) non-unital dynamics implemented by \textit{spontaneous decay}. 

In some cases, it can be beneficial to express the quantum dynamics in Eq.~\eqref{eq:kraus_decomposition} in terms of an equation of motion for $\rho$, which is always possible if the dynamics are Markovian (and time homogeneous). In that scenario, the quantum evolution can be written in exponential form as $\mathcal{E}(t) = \exp(\mathcal{L} t)$ where $\mathcal{L}$ is the so-called Lindbladian. Then, the equation of motion for $\rho$ yields simply $\partial_t\rho = \mathcal{L}[\rho]$, and it can be cast in the form of the well-known Gorini-Kossakowski-Sudarshan-Lindblad equation ($\hbar=1$) \cite{GoKoSu1976,Li1976}
\begin{equation}
\mathcal{L}[\rho] = -i[\text{H},\rho] + \sum_k \kappa_k \left( \text{J}_k \rho \text{J}_k^\dagger -\frac{1}{2}\{\text{J}_k^\dagger \text{J}_k,\rho\}\right),\label{eq:gksl_equation}
\end{equation}
where $\text{H}$ is the Hamiltonian, $\kappa_k\geq 0$ are the dissipation rates, and $\text{J}_k$ the jump operators. Without loss of generality, we take $\text{H} = \omega \sigma_z/2$, where $\sigma_z$ is the z-Pauli operator, and label its eigenstates according to $\sigma_z\ket{0} = \ket{0}, ~ \sigma_z\ket{1} = - \ket{1}$. Also, we specify the measurement $\text{M}$ to have the eigenstates $\ket{m_+}=\cos(\theta/2)\ket{0} + \sin(\theta/2)\ket{1}$ and $\ket{m_-} = -\sin(\theta/2)\ket{0} + \cos(\theta/2)\ket{1}$, where $\theta\in[0,\pi]$, and we chose the target state $\ket{m_\star} = \ket{m_+}$. 

We start by studying the evolution under unitary dynamics, which is recovered from Eq.~\eqref{eq:gksl_equation} for vanishing dissipation rates $\kappa_k=0$. For $N=2$, the matrix $\mathbf{W}(s)$ has a single entry which simplifies substantially the computation. It is convenient to introduce the time-averaged conditional probabilities $a(i|j) = \int_0^\infty dt \varphi(t) p(i,t|j)$. Then, the direct application of Eqs.~\eqref{eq:tstar}--\eqref{eq:tis} brings to

\begin{align}
    &T_- = \frac{\tau}{a(+|-)},\label{eq:t-_qubit_unitary}\\
    &T_+ = T_\star = \tau+a(-|+) T_- = 2\tau,\label{eq:t+_qubit_unitary}
\end{align}

where we have used that $a(+|-) = a(-|+)$ under unitary dynamics (and $N=2$). We remark that Eqs.~\eqref{eq:t-_qubit_unitary}--\eqref{eq:t+_qubit_unitary} are not valid for $\theta = 0,\pi$. These two angles lead to a quantum dynamics with block structure and should be treated separately according to the discussion above Eq.~\eqref{eq:NcTau}.

The results of the numerical simulation are shown for an exponential probability density function $\varphi_{\text{exp}}(t) = \tau^{-1}\exp(-t/\tau)$, for which we obtain 
\begin{equation}
    T_- = \frac{2\tau}{\sin^2(\theta)}\left(1+\frac{1}{(\omega\tau)^2}\right).\label{eq:t-_qubit_exp}
\end{equation}
Equation~\eqref{eq:t-_qubit_exp} is simple but it reveals an intriguing possibility. For $\varphi(t)=\varphi_{\text{exp}}(t)$, the switching time admits an optimal value of the mean measurement time $\tau = \omega^{-1}$ for which $T_-$ is minimum. This is due to the combination of two different phenomena: on one hand, for small $\tau$ we have quantum Zeno effect \cite{MiSu77} and the system is unable to leave the initial state; on the other hand, for a large value of $\tau$ the system is seldom measured, which makes it unoptimal. In \cite{FrKeBa17} this is studied in detail for stroboscopic measurements. The optimallity of the resetting is also found for systems where the resetting is of probabilistic nature instead of quantum \cite{EvMa11,EvMa11p,MoVi13,ChSc15,CaMe15,PaKu16,NaGu16,MoMa17,Sh17,MaCaMe19,KuGu19,Ma19,BoChSo19,SaInAj19,MaMo19,Si20,EvMa19, MaCaMe19_1, PaKuRe19,ReUrKl14,RoReUr15,Be18,PaRe17,Reuveni16,ChSo18,DeRaRe20}. It is worth mentioning that, contrarily to the mean switching time, Eq.\eqref{eq:t+_qubit_unitary} does not admit an optimal value for $\tau$.

In Fig.~\ref{fig:all}a, we compare the analytic formula in Eq.~\eqref{eq:t+_qubit_unitary} and Eq.~\eqref{eq:t-_qubit_exp} with the results of the numerical simulation. Both the mean switching time $T_-$ and mean return time $T_\star = T_+$ show an excellent agreement. Further numerical investigation reveals that the analytical predictions remain accurate for other normalizable $\varphi(t)$ (e.g., normally distributed,...). 

The second example consists in studying the unital dynamics under dephasing noise. Such an evolution is described by the following Lindbladian:
\begin{equation}
    \mathcal{L}[\rho] = -i\frac{\omega}{2}[\sigma_z,\rho] + \kappa (\sigma_z \rho \sigma_z - \rho), 
\end{equation}
with a single dissipation rate $\kappa_1=\kappa > 0$ and a single jump operator $\text{J}_1 = \sigma_z$. For $\kappa=0$ one would recover unitary dynamics. The direct application of  Eqs.~\eqref{eq:tstar}--\eqref{eq:tis} leads to the same expressions for $T_+$ and $T_-$ as in Eqs.~\eqref{eq:t-_qubit_unitary}--\eqref{eq:t+_qubit_unitary}, showing the universal behavior $T_\star = T_+ = 2\tau$. Naturally, the numerical value of the averaged conditional probabilities $a(i|j)$ is in this case different. This is explicitly observed by taking $\varphi(t) = \varphi_\text{exp}(t)$ which leads to
\begin{equation}
    T_- = \frac{2\tau}{\sin ^2(\theta)}\frac{\left((2 \kappa  \tau +1)^2+ (\omega\tau)^2 \right)}{(2\kappa
   \tau)^2 +2\kappa\tau +(\omega\tau)^2},\label{eq:t-_qubit_unital_exp}
\end{equation}
which reduces to Eq.~\eqref{eq:t-_qubit_exp} for $\kappa \ll \omega$. The result of the numerical simulation is compared to Eq.~\eqref{eq:t-_qubit_unital_exp} and displayed in Fig.~\ref{fig:all}a for different increasing values of $\kappa$ showing, again, a very good agreement. 

Lastly, we consider the case of non-unital dynamics in the presence of spontaneous decay. This evolution is described by the Lindbladian:
\begin{equation}
    \mathcal{L}[\rho] = -i\frac{\omega}{2}[\sigma_z,\rho] + \kappa (\sigma_- \rho \sigma_-^\dagger - \frac{1}{2}\{\sigma_-^\dagger \sigma_-,\rho\}), 
\end{equation}
with a single dissipation rate $\kappa_1 = \kappa > 0$ and a single jump operator $\text{J}_1 = \sigma_- = \ketbra{1}{0}$. For the non-unital dynamics under scrutiny, it is no longer true that $a(+|-) = a(-|+)$. Then, using of  Eqs.~\eqref{eq:tstar}--\eqref{eq:tis} leads to

\begin{align}
    &T_- = \frac{\tau}{a(+|-)}\label{eq:t-_qubit_decay},\\
    &T_+ = T_\star = \tau\left(1+\frac{a(-|+)}{a(+|-)}\right).\label{eq:t+_qubit_decay}
\end{align}

The ratio in Eq.~\eqref{eq:t+_qubit_decay} is a measure of the asymmetry in the averaged conditional probabilities $a(i|j)$. This asymmetry is the underlying reason behind the breaking of the universal behavior in Eq.~\eqref{eq:Ntau}. Again, we can proceed further by setting $\varphi(t) = \varphi_\text{exp}(t)$, which leads to cumbersome expressions for $T_+$ and $T_-$ that the interested reader can find in Appendix \ref{appendix:qubit_decay}.

In Fig.~\ref{fig:all}b we compare the analytical result of $T_+$ with the result of the numerical simulation for some values of $\kappa$. Special limits are $\kappa \ll \omega$ and $\kappa \gg \omega$. The former is the unitary limit, where we see that a larger region of $\theta$ becomes closer to $2 \tau$, as expected. In the latter limit, a large $\kappa$ drives a rapid decay from $\ket{0}$ to $\ket{1}$. This is captured as a divergence of $T_+$ as $\theta \to 0$ and $T_+ \to \tau$ as $\theta \to \pi$. A similar analysis can be done in Fig.~\ref{fig:all}c with $T_-$ instead of $T_+$. 

In Appendix \ref{appendix:qutrit_unitary} we provide an analysis of the return and the switching times for the qutrit (N=3) case under unitary evolution.

\section{Conclusions}

We have modelled a resetting process using a built-in feature of quantum theory: the collapse of the wave function. Within the framework of renewal equations, we have computed the mean switching and return times of any type of quantum evolution and measurement time pdf. Their expressions are found using the experimentally accessible data encoded in the transition probabilities. The mean return time shows a universal behavior for all unital evolutions. This result is particularly interesting for experimental noisy systems. Such universal behavior is only broken due to the asymmetry between measurement states introduced by non-unital processes. Moreover, in some scenarios, the reset rate can be optimized to achieve the fastest possible switching time between states.

An interesting direction of research would be that of generalizing the measurement to possibly degenerate outputs. Such a process would keep the coherence within the degenerate subspaces leading to potentially different universal behaviors and optimal reset rates. Hence, a more general study of the measurement-induced resetting in open quantum systems is a promising venue of research.

\begin{acknowledgments}
\textit{Acknowledgments.}-- We thank John Calsamiglia, Ramon Muñoz-Tapia, Vicenç Méndez and Albert Sanglas for enlightening discussions. We also thank Eli Barkai for his insightful comments. ARC acknowledges financial support from the Spanish MINECO/ AEI FIS2016-80681-P, PID2019-107609GB-I00, from the Catalan Government: projects CIRIT 2017-SGR-1127, AGAUR FI-2018-B01134, and QuantumCAT 001-P-001644 (RIS3CAT comunitats), co-financed by the European Regional Development Fund (FEDER). JO acknowledges financial support from the grants FPA2017-88915-P and SEV-2016-0588 from MINECO. IFAE is partially funded by the CERCA program of the Generalitat de Catalunya. AMP acknowledges financial support from the Ministerio de Economia y Competitividad through Grant No.
CGL2016-78156-C2-2-R.\\

All authors have contributed equally to this work.
\end{acknowledgments}

\appendix

\onecolumngrid

\section{Derivation of Eqs. (7)--(8) in the main text}
\label{appendix:derivation_of_eqs_7_and_8}

It is possible to write renewal equations for the survival probabilities starting from the different states in the measurement basis as:
\begin{align}
&Q_\star(t) = \bar{\varphi}(t) + \sum_{j\neq\star} \int_0^t dt' \varphi(t') p(j,t'|\star) Q_j(t-t')dt',\label{eq:renewalstar_app}\\
&Q_i(t) = \bar{\varphi}(t) + \sum_{j\neq\star}\int_0^t dt' \varphi(t') p(j,t'|i) Q_j(t-t')dt',\label{eq:renewal_app}
\end{align}
where $\bar{\varphi}(t)=\int_t^\infty \varphi(t')dt'$ is the survival probability of the measurement time distribution. Performing the Laplace transform at both sides of the Eq.~\eqref{eq:renewalstar_app} and Eqs.~\eqref{eq:renewal_app} and using the convolution theorem, we get
\begin{align}
Q_\star (s)= \hat{\bar{\varphi}}(s) +\sum_{j\neq\star} \mathbb{L}[\varphi(t)p(j,t|\star)](s) Q_j (s),\label{eq:survival_star_app}\\
Q_i (s) = \hat{\bar{\varphi}}(s) + \sum_{j\neq\star} \mathbb{L}[\varphi(t)p(j,t|i)](s) Q_j (s).\label{eq:survival_multi_app}
\end{align}

We note that Eqs.~\eqref{eq:survival_multi_app} are self-contained in the subspace orthogonal to $\ket{m_\star}$. Eq.~\eqref{eq:survival_star_app} and Eq.~\eqref{eq:survival_multi_app} can be compactly written by introducing the $(N-1)$-dimensional vector $\mathbf{w}_{\star,i}(s) = \mathbb{L}[\varphi(t)p(i,t|\star)](s)$ and the matrix $(N-1)\times(N-1)$-dimensional $\mathbf{W}_{ij}(s) = \mathbb{L}[\varphi(t) p(i,t|j)](s)$. Also, we take the survival probabilities $Q_i(s)$ as the components of a vector $\mathbf{Q}(s)$. In this way, straightforward algebra leads to the vector equations
\begin{align}
&\hat{Q}_\star(s) = \hat{\bar{\varphi}}(s)\left(1+ \mathbf{w}_{\star}^T\left(\mathbf{1}-\mathbf{W}^T(s)\right)^{-1}\mathbf{v}_1\right),\\
&\hat{\mathbf{Q}}(s) = \hat{\bar{\varphi}}(s)(1-\mathbf{W}^T(s))^{-1}\mathbf{v}_1,
\end{align}
where $\mathbf{v}_1 = (1,\cdots,1)^T$. 

The mean first detection time $T_i$ is related to its associated survival probability $Q_i(t)$ through $T_i = \hat{Q}_i(s=0)$. Therefore, the mean first detection time is found
\begin{align}
&T_\star = \tau \left(1+ \mathbf{w}_{\star}^T\left(\mathbf{1}-\mathbf{W}^T(s=0)\right)^{-1}\mathbf{v}_1\right)\label{eq:equations_tstar_app},\\
&\mathbf{T} = \tau(1-\mathbf{W}^T(s=0))^{-1}\mathbf{v}_1,\label{eq:equations_ti_app}
\end{align}
where we have used $\tau=\hat{\varphi}(s=0)$.

\section{Derivation of Eq. (9) in the main text}
\label{appendix:derivation_of_eq_9}

Let $\text{M}$ be a non-degenerate quantum measurement defined by the set of projectors $\{\ketbra{m_i}{m_i}\}$. On the one hand, the transition probabilities $p(i,t|j) = \langle m_i|\mathcal{E}(t)[\ketbra{m_j}{m_j}]|m_i\rangle$ obviously fulfill that $\sum_i p(i,t|j) = 1$ for any evolution $\mathcal{E}(t)$. On the other hand, summing over the conditions $\sum_j p(i,t|j) = \langle m_i|\mathcal{E}(t)[\mathbf{1}]|m_i \rangle$ is in general different from one. However, \textit{if} the evolution is unital we also find $\sum_j p(i,t|j) = 1$.

Again, we consider the matrix $\mathbf{W}_{ij}(s) = \mathbb{L}[\varphi(t) p(i,t|j)](s)$ for $i,j \neq \star$ and, similarly, the vector $\mathbf{w}_{\star,i}(s) = \mathbb{L}[\varphi(t) p(i,t|\star)](s)$. For a unital quantum evolution $\mathcal{E}(t)$, we find the relation
\begin{align}
    \sum_j \mathbf{W}_{ij}(s) &= \sum_{j \neq \star} \mathbb{L}[\varphi(t)( 1-p(i,t|\star))](s)\nonumber\\
    &= \hat{\varphi}(s) - \mathbf{w}_{\star,i}(s). \label{eq:completeness_app}
\end{align}
Equivalently, we can write down the matrix form of Eq.~\eqref{eq:completeness_app} as
\begin{align}
\mathbf{w}_\star^T(s) = \mathbf{v}_1^T \left(\hat{\varphi}(s)-\mathbf{W}^T(s)\right),\label{eq:epic_relation_general}
\end{align}
where $\mathbf{v}_1 = (1,\cdots,1)$ is a vector of length $N-1$.

\section{Universal behavior under a quantum evolution $\mathcal{E}(t)$ with block structure}
\label{appendix:universal_behavior_under_block_structure}

In this appendix, we discuss in more detail the form of the universal behavior $T_\star = N\tau$ when the quantum evolution $\mathcal{E}(t)$ is unital and has a block structure.
We start considering a general quantum evolution $\mathcal{E}(t)$ and add an extra label $\mu = c,d$ to the measurement states $\ket{m_i,\mu}$ to denote whether they are dynamically connected or disconnected from the \textit{target} state. More precisely, a disconnected state $\ket{m_i,d}$ fulfills
\begin{align}
    p(i,t|\star) = \langle m_i,d| \mathcal{E}(t)[\ketbra{m_\star,c}{m_\star,c}] |m_i,d\rangle = 0 \qquad \forall t,\label{eq:direct_disconnected}
\end{align}
while the a connected state is a state that does not fulfill the above condition. Obviously, the target state is always connected to itself since at time $t=0$ the quantum evolution has to  be equal to the identity map $\mathcal{E}(0) = \mathcal{I}$. In general, it can happen that for a disconnected state $\ket{m_i,d}$ 
\begin{align}
        p(\star,t|i) = \langle m_\star,c| \mathcal{E}(t)[\ketbra{m_i,d}{m_i,d}] |m_\star,d\rangle \neq 0.\label{eq:reverse_disconnected}
\end{align}
Even though Eq.~\eqref{eq:reverse_disconnected} may seem unintuitive at first glance, there are well-known examples of it. For instance, if we measure a two level atom with spontaneous decay in its energy eigenbasis the ground state is disconnected from the excited state while the converse is not true. 

If we now assume $\mathcal{E}(t)$ to be unital, the possibility stated by Eq.~\eqref{eq:reverse_disconnected} is ruled out. This can be seen as the consequence of two facts:
\begin{enumerate}
    \item Any unital evolution can be written as a linear-affine combination (see \cite{MeWo2009}): 
    \begin{align}
\mathcal{E}(t) = \sum_l \lambda_l \text{U}_l(t) \rho \text{U}_l(t)^\dagger.\label{eq:unital_general_form}
\end{align}
    \item Any unitary $\text{U}_l(t)$ can be written as $\text{U}_l(t) = \exp(-i \text{A}_l(t))$  where $\text{A}_l(t)$ is Hermitian. Hence, $\ket{m_i,d}$ is disconnected from $\ket{m_\star,c}$ only if for all times $t$ and positive integers $n$ 
    \begin{align}
        \langle m_i,d |\text{A}^n_l(t)|m_\star,c \rangle = \langle m_\star,c |\text{A}^n_l(t)|m_i,d \rangle= 0,
    \end{align}
    which implies that $\text{A}_l(t)$ (and consequently $\text{U}_l(t)$) has a block structure. 
\end{enumerate}

Then, denoting $\Pi_c$ and $\Pi_d$ the projectors $\Pi_\mu = \sum_i \ketbra{m_i,\mu}{m_i,\mu}$, the unital quantum evolution $\mathcal{E}(t)$ has a block structure
\begin{align}
\Pi_\nu \mathcal{E}(t)[\Pi_\mu] \propto \delta_{\mu\nu}. \label{eq:block_map}
\end{align}
Together with the unital property, Eq.~\eqref{eq:block_map} implies that the quantum evolution is unital block-wise, i.e.  $\mathcal{E}(t)[\Pi_\mu] = \Pi_\mu$. Therefore, one could repeat the same derivation of the main text, where only the subset of connected states are considered. Then, one would arrive to 
\begin{align}
    T_\star = N_c \tau.
\end{align}
where $N_c = \text{tr}[\Pi_c]$ is the number of connected states. Two final remarks are in order: (i) if the disconnected space is zero-dimensional $N_c = N$ and we recover $T_\star = N\tau$, and (ii) the block structure of $\mathcal{E}(t)$ makes the derivation of Eq.~(10) not valid since the matrix $\mathbf{1}-\mathbf{W}^T(0)$ is not invertible in that scenario.

\ignore{
\Andreu{Ref: Unital Quantum Channels – Convex Structure and Revivals of Birkhoff’s Theorem}

For a single unitary of the form $\exp(-i H t)$ where $H = H^\dagger$, the only way that $\ket{i|U|j}=0\forall t$ is that $H$ has a block structure and then this implies that U has a block structure too. For the combination of unitaries this should also hold!

Any unital quantum evolution $\mathcal{E}(t)$ is an linear-affine combination of unitary maps, i.e. it can be decomposed as 

for $\lambda_i \in \mathbb{R}$, and $\sum_i \lambda_i = 1$ and $\text{U}_i(t)$ unitary.

Then, one can proof that the block structure is the only possible unital disconnected dynamics. More formally that given a unital map in the form Eq.~\eqref{eq:unital_general_form}
\begin{align}
    \text{tr}[\Pi_d \mathcal{E}(t)[\Pi_c]]=0 \iff \text{tr}[\Pi_c \mathcal{E}(t)[\Pi_d]]=0
\end{align}
\textit{Proof.}--- It is easy to proof that,
\begin{align}
    0 =& \text{tr}[\Pi_d \mathcal{E}(t)[\Pi_c]] =\sum_i \lambda_i \text{tr}[\Pi_d \text{U}_i(t) \Pi_c \text{U}_i(t)^\dagger]\\
    =& \left(\sum_i \lambda_i \text{tr}[\Pi_d \text{U}_i(t) \Pi_c \text{U}_i(t)^\dagger]\right)^\dagger = \text{tr}[\Pi_c \mathcal{E}(t)[\Pi_d]],
\end{align}}

\section{The qubit ($N=2$): decay dynamics}
\label{appendix:qubit_decay}

For a qubit under decay dynamics (i.e., under the Lindbladian in Eq.~(18) of the main text), the cumbersome expressions for the mean switching time $T_-$ and the mean first return time $T_+$ for $\varphi(t) = \varphi_\text{exp}(t)$ yield
\begin{align}
    &T_- = \tau \frac{(\kappa  \tau +1) \csc ^2\left(\frac{\theta }{2}\right) \left((\kappa  \tau +2)^2+
   (2\omega\tau)^2 \right)}{(\kappa  \tau +1) \left(\kappa\tau  (\kappa  \tau +2)+(2\omega\tau) ^2\right)-\cos (\theta ) \left(\kappa  (\kappa \tau +2)-(2\omega\tau
   )^2\right)},\label{eq:t-_qubit_decay_exp}\\
    &T_+ = \tau  \left(1+\frac{\cot ^2\left(\frac{\theta }{2}\right) \left(\cos (\theta ) \left(\kappa\tau
    (\kappa  \tau +2)-(2\omega\tau)^2\right)+(\kappa  \tau +1) \left(\kappa\tau  (\kappa 
   \tau +2)+(2 \omega \tau)^2\right)\right)}{(\kappa \tau +1) \left(\kappa \tau (\kappa 
   \tau +2)+ (2\omega\tau)^2 \right)-\cos (\theta ) \left(\kappa\tau (\kappa \tau +2)-(2
  \omega\tau)^2\right)}\right),\label{eq:t+_qubit_decay_exp}
\end{align}
which reduce to the expressions of the unitary case as $\kappa/\omega \to 0$.

\section{The qutrit ($N = 3$): unitary dynamics}
\label{appendix:qutrit_unitary}

We consider a $N = 3$ system evolving under the unitary dynamics $U(t) = e^{-i \omega t S_x}$, where $S_x$ is a representation of the $x$ component of the spin-1 operator. The measurement basis is chosen to be the eigenbasis of the $S_z$ operator, which we label by $|+1\rangle, |0\rangle, |-1\rangle$ according to their eigenvalues. There are nine mean times, which we will label as $T_{\text{in} ~ \text{out}}$. Whenever $|\text{in}\rangle = |\text{out}\rangle$ we will have a mean first return time, $T_\star$, and in the other cases we will have mean switching times.

Mean switching times are computed using Eq.(8) in the main text, while the remaining ones are simply $3 \tau$. We choose $\varphi(t) = \varphi_\text{exp}(t)$ to perform the calculation. We show one particular example in detail and then state the final results.

Taking $\ket{m_\star} = \ket{+1}$, the matrix $\mathbf{W}$ is
\begin{equation}
    \mathbf{W} = \begin{pmatrix}
\mathbb{L}\left[ \varphi(t) p(0, t | 0)\right](s) & \mathbb{L}\left[ \varphi(t) p(0, t | -)\right](s) \\
\mathbb{L}\left[ \varphi(t) p(-, t | 0)\right](s) & \mathbb{L}\left[ \varphi(t) p(-, t | -)\right](s)
\end{pmatrix},
\end{equation}
and so the mean switching times are
\begin{equation}
\begin{pmatrix}
T_{0 +} \\
T_{- +}
\end{pmatrix}
= \tau
\left[ \mathbf{1} -
 \begin{pmatrix}
\mathbb{L}\left[ \varphi(t) p(0, t | 0)\right](s=0) & \mathbb{L}\left[ \varphi(t) p(-, t | 0)\right](s=0) \\
\mathbb{L}\left[ \varphi(t) p(0, t | -)\right](s=0) & \mathbb{L}\left[ \varphi(t) p(-, t | -)\right](s=0)
\end{pmatrix}
\right]^{-1}
\begin{pmatrix}
1 \\
1
\end{pmatrix}.
\end{equation}
The transition probabilities are computed using $p(i,t|j) = |\bra{i} e^{-i \omega t S_x} \ket{j}|^2$, and so the elements inside the matrix $\mathbf{W}(s=0)$ explicitly become $\mathbf{W}_{ij}(s=0) = \int_0^\infty ~ dt \exp(-t/\tau) |\bra{i} e^{-i \omega t S_x} \ket{j}|^2 / \tau$.

Carrying out the calculation in full yields the final result for $\ket{m_\star} = \ket{+}$, which we find to be
\begin{align}
    T_{0+} &= \tau \left( \frac{7}{2} + \frac{2}{(\omega \tau)^2}\right) \\
    T_{-+} &= 3\tau \left( 1 + \frac{1}{(\omega \tau)^2}\right).
\end{align}
In an analogous way, the results for $\ket{m_\star} = \ket{0}$ are
\begin{align}
    T_{+0} &= \tau \left( 4 + \frac{1}{(\omega \tau)^2}\right) \\
    T_{-0} &= T_{+0}.
\end{align}
Finally, for $\ket{m_\star} = \ket{-1}$ we find
\begin{align}
    T_{+-} &= T_{-+} \\
    T_{0-} &= T_{0+}.
\end{align}
The analytic three mean return times and the six mean switching times are compared with the results of the numerical simulation in Table~\ref{tab:n3}. 

\begin{table}[h]
\centering
\begin{tabular}{| l | l | l | l | l | l | l | l | l | l |}
\hline
$T_{\text{in}~\text{out}}$ & $T_{++}$ & $T_{0+}$ & $T_{-+}$ & $T_{+0}$ & $T_{00}$ & $T_{-0}$ & $T_{+-}$ & $T_{0-}$ & $T_{--}$ \\
\hline
Theory & 3 & 5.5 & 6 & 5 & 3 & 5 & 6 & 5.5 & 3 \\
Simulation & 3.06 & 5.56 & 5.97 & 5.02 & 3.02 & 5.02 & 5.94 & 5.58 & 3.02 \\
\hline
\end{tabular}
\caption{Comparison between the results from the numerical simulations and the analytical formulas for the qutrit ($N=3$). We have taken units $\omega = \tau^{-1} = 1$ and considered the probability density function $\varphi_\text{exp}(t)$. Each process has been simulated 10,000 times.}
\label{tab:n3}
\end{table}

\bibliographystyle{unsrtnat}
\bibliography{QuantumResetting}

\end{document}